# Experiments on thermoelectric properties of quantum dots

**Artis Svilans, Martin Leijnse and Heiner Linke**

NanoLund and Solid State Physics, Lund University, Box 118, 221 00 Lund, Sweden

*E-mail: artis.svilans@ftf.lth.se, heiner.linke@ftf.lth.se

## Abstract

Quantum dots (QDs) are good model systems for fundamental studies of mesoscopic transport phenomena using thermoelectric effects because of their small size, electrostatically tunable properties and thermoelectric response characteristics that are very sensitive to small thermal biases. Here we provide a review of experimental studies on thermoelectric properties of single QDs realized in two-dimensional electron gases, single-walled carbon nanotubes and semiconductor nanowires. A key requirement for such experiments is methods for nanoscale thermal biasing. We briefly review the main techniques used in the field, namely, heating of the QD contacts, side heating and top heating, and touch upon their relative advantages. The thermoelectric response of a QD as a function of gate potential has a characteristic oscillatory behavior with the same period as is observed for conductance peaks. Much of the existing literature focuses on the agreement between experiments and theory, particularly for amplitude and line-shape of the thermovoltage $V_{th}$. A general observation is that the widely used single-electron tunneling approximation for QDs has limited success in reproducing measured $V_{th}$. Landauer-type calculations are often found to describe measurement results better despite the large electron-electron interactions in QDs. More recently, nonlinear thermoelectric effects have moved into the focus of attention and we offer a brief overview of experiments done so far. We conclude by discussing open questions and avenues for future work, including the role of asymmetries in tunnel- and capacitive couplings in thermoelectric behavior of QDs.

# 1. Introduction

Quantum dots (QDs) in electrical circuits are sub-micrometer sized metallic or semiconducting regions with an electrochemical potential that varies rapidly with the number of occupying charge carriers. Electron transport through QDs has been studied for several decades, both out of fundamental interest in classical and quantum electron transport phenomena at the nanoscale [1] and as potentially useful building blocks in novel electronic circuits [2]. For the purpose of this review we use the QD concept both for devices obeying classical and quantum Coulomb blockade.

A widely studied parameter of QDs is the two-terminal conductance, $G$, which at low enough temperature becomes strongly nonlinear with respect to application of electrical bias. This characteristic is a manifestation of the discrete nature of QD's electrochemical potential spectrum. This review focuses on the thermoelectric properties of QDs, which have been investigated much less than the conductance.

The number of experimental studies on QD thermoelectric properties is limited. Most of the existing work has focused on the Seebeck effect [3], which manifests itself as a potential difference $V_{th}$ that develops in response to a temperature difference $\Delta T$ across a QD in an open-circuit configuration. It has been demonstrated that $V_{th}$ is more sensitive to certain transport phenomena in QDs than $G$ [4, 5, 6] already in the linear response regime and generally contains complementary information about the device. For example, $V_{th}$ signals can remain strong even when $G$ becomes too small to be precisely measured, as it is the case when transport is blocked by Coulomb blockade. In addition, QDs have been predicted to be efficient thermoelectric convertors because of their ability to selectively transport thermally excited carriers [7, 8, 9, 10]. We focus this review on experiments on single QDs and refer to [11] for a review on experiments on double-QDs.

Measurements of $V_{th}$ in QDs were among the first thermoelectric transport experiments carried out on mesoscopic devices [12, 13]. Despite these first results showing good agreement with theoretical models, it has generally been difficult to achieve quantitative agreement between theoretical models and thermoelectric measurements. This is somewhat surprising because the same theoretical models describing $G$ behavior typically have been able to explain measurement data well.

In comparison to plain conductance characterization [14], thermoelectric characterization requires additional measurements of the thermal bias $\Delta T$ applied to a device which are experimentally challenging.. Furthermore, conductance and thermoelectric measurements together contain more information about a device and theoretical models might need to be more detailed to capture all the relevant physics.

In the following we start with a brief overview of the key concepts in experiments on QD-based thermoelectric devices. We then provide an intuitive, physical picture of the characteristic, oscillatory thermoelectric response of QDs, followed by a review of experimental thermoelectric studies on QDs. Finally, we discuss open or little discussed aspects of the field.

## 2. Devices and methods

A characteristic property of a QD is its strongly nonlinear conductance. The underlying reason for the nonlinearity is the electrochemical potential of a QD that varies significantly with the number of charge carriers (typically electrons) occupying it [1, 15]. This can be thought of as a set of quasi-discrete energies (resonances) via which electrons can be transported through the QD.

QDs in electrical circuits are defined and probed using tunnel-junctions, which are thin, controlled potential barriers for electrons. Two tunnel-junctions can define small regions in which charging and quantum confinement effects [16, 17] become important, resulting in the discretization of the electrochemical potential of the QD. At the same time tunnel-junctions maintain some transparency for electron tunneling that allows probing the resulting resonance spectrum of the QD. Energy separation between resonances, $\Delta\varepsilon$, generally increases with shrinking dimensions of the QD.

In an electric or thermoelectric measurement setup, the QD is tunnel-coupled to two relatively big electrodes that can be thought of as electron reservoirs and are characterized by their electrochemical potentials, $\mu_L$ and $\mu_R$, and temperatures, $T_L$ and $T_R$ (Fig. 1). In the simplest approximation, tunnel-coupling can be characterized by constant tunnelling rates, $\Gamma_L$ and $\Gamma_R$. In addition, the QD is capacitively coupled to the surrounding electromagnetic environment. In a typical device, the capacitive coupling of a QD can be reduced to two couplings to the reservoirs, $C_L$ and $C_R$, plus a coupling to an additional gate electrode, $C_G$. Varying electrical potential of the gate electrode allows great flexibility as it has an effect of shifting the resonances in energy relative to $\mu_L$ and $\mu_R$ which can be set externally.

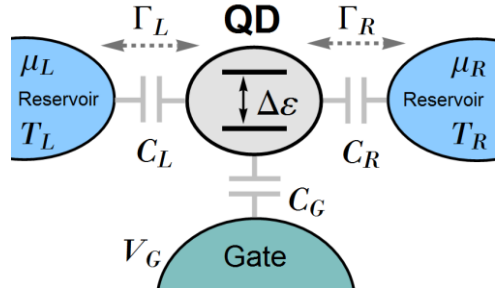

Figure 1: Schematic view of a QD layout. A QD (object in gray), is characterized by a quasi-discrete resonance spectrum with typical resonance separation $\Delta\varepsilon$. The QD is capacitively coupled to a gate (in green) with two-terminal capacitance $C_G$ and to two electron reservoirs (in blue) with two-terminal capacitances $C_L$ and $C_R$ to the left (label L) and right (label R) reservoirs, respectively. Reservoirs are characterized by their temperatures $T_L$ and $T_R$, and their electrochemical potentials $\mu_L$ and $\mu_R$. Electrical potential of the gate electrode is $V_G$. Electron exchange between reservoirs and a QD takes place by tunneling with characteristic frequencies (tunneling rates) $\Gamma_L$ and $\Gamma_R$.

**Relevant energy scales**

The physical behavior of QDs is strongly determined by the relative size of a number of energy scales that we briefly discuss in the following.

**Charging.** One reason for discretization of the QD spectrum in resonances is the electrostatic interaction between an added charge and other charges already occupying the QD. In typical devices reviewed here, the strength of this interaction, often called charging energy and denoted $E_C$, is a few meV (corresponding to $k_B T$ at a few tens of Kelvin). It is common expressed as $E_C = e^2/(2C)$, where $e$ is the elementary charge and $C \approx C_G + C_L + C_R$ is the self-capacitance of a QD, which scales roughly linearly with QD size.

**Quantum confinement.** Discrete single-particle states, or QD orbitals, are separated by quantum-confinement energy $\Delta E$, which increases roughly quadratically with decreasing QD size and therefore generally can vary in a wider range than $E_C$. In the experiments reviewed here, $\Delta E$ is typically of the same order of magnitude or smaller than $E_C$. In the constant-interaction model (CIM), the lowest total $\Delta\varepsilon$ is thought as approximately a sum of the two contributions, $E_C$ and $2\Delta E$ [18]. Additional resonances can appear because of excited states, where one or more electrons occupy higher-energy orbitals rather than simply filling the lowest available ones.

**Tunnel-coupling.** The inverse lifetime of an electron on the QD scales with $\Gamma = \Gamma_L + \Gamma_R$. It is also a measure of tunnel-coupling strength between the QD and the rest of the circuit. As given by Heisenberg's uncertainty principle, this finite lifetime is related to a broadening $\hbar\Gamma$ of resonance energies, with $\hbar$ being the reduced Plank's constant. If one neglects spin effects, the maximal current through a single QD resonance is given by $I_{max} = \pm e\Gamma_L\Gamma_R/\Gamma = \pm e\gamma$ [19]. $I_{max}$, for a given $\Gamma$, reaches maximum when $\Gamma_L = \Gamma_R$. Assuming a typical value of $I_{max} = 1$ nA then gives $\Gamma = 4\gamma = 25.0$ GHz, resulting in $\hbar\Gamma = 16.4$ μeV (corresponding to $k_BT$ at 191 mK). Whereas this is a relatively small amount of energy, in case $\Gamma_L$ and $\Gamma_R$ are very different (asymmetric), the numerical factor between $\Gamma$ and $\gamma$ can be much bigger than 4.

**Temperature.** QDs are operated outside of thermal equilibrium and therefore it is difficult to define their temperature. Electron reservoirs, on the other hand, can often be assumed to be near thermal equilibrium, in which case electrons occupy states according to Fermi-Dirac statistics with a characteristic energy scale of $k_BT$. In transport measurements, the physics of QDs is best resolved when $k_BT$ and $k_B\Delta T$ are both much smaller than $\Delta\varepsilon$. Due to experimental limitations of increasing $\Delta\varepsilon$, this typically means that experiments are done at temperatures below 10 Kelvin or even below 100 mK.

## Experimental systems

During the past few decades, several convenient ways have been found to integrate well-defined QDs into electrical circuits. Here we give a brief introduction to the types of QD devices used for thermoelectric measurements and discussed in Sec. 4.

The first experiments on thermoelectric properties of QDs were carried out on QDs defined in two-dimensional electron gases (2DEGs). Essentially they are two-dimensional layers of highly mobile electrons formed at the interface of two semiconductors, typically modulation-doped GaAs/AlGaAs [20]. Electron channels can be defined either by etching, or, as it is more often done in this system, by electrostatic gates deposited on top, at a small distance from the 2DEG (Fig. 2a). In the latter case tunnel-junctions are created by a saddle-type potential formed below a pair of near-by gate electrodes (so-called split gates). The parts of the 2DEG leading to tunnel-junctions serve as electron reservoirs, and the part of 2DEG between the tunnel-junctions as a QD.

An alternative approach is to define QDs in quasi-one-dimensional structures, such as nanotubes or nanowires (Fig. 2b and 2c, respectively). For example, Schottky barriers at the interface between metal leads and nanowires or carbon nanotubes can play the role of tunnel-junctions [21, 22]. In this case, metal leads become electron reservoirs. Alternatively, a QD can be embedded directly in a heterostructured semiconductor nanowire (e.g. InAs), by using higher bandgap (e.g. InP) segments as tunnel-junctions (Fig. 2d) [23, 24]. Ideally, the metallic leads then have Ohmic contacts to the nanowire and the parts of nanowire approaching a QD are then the reservoirs.

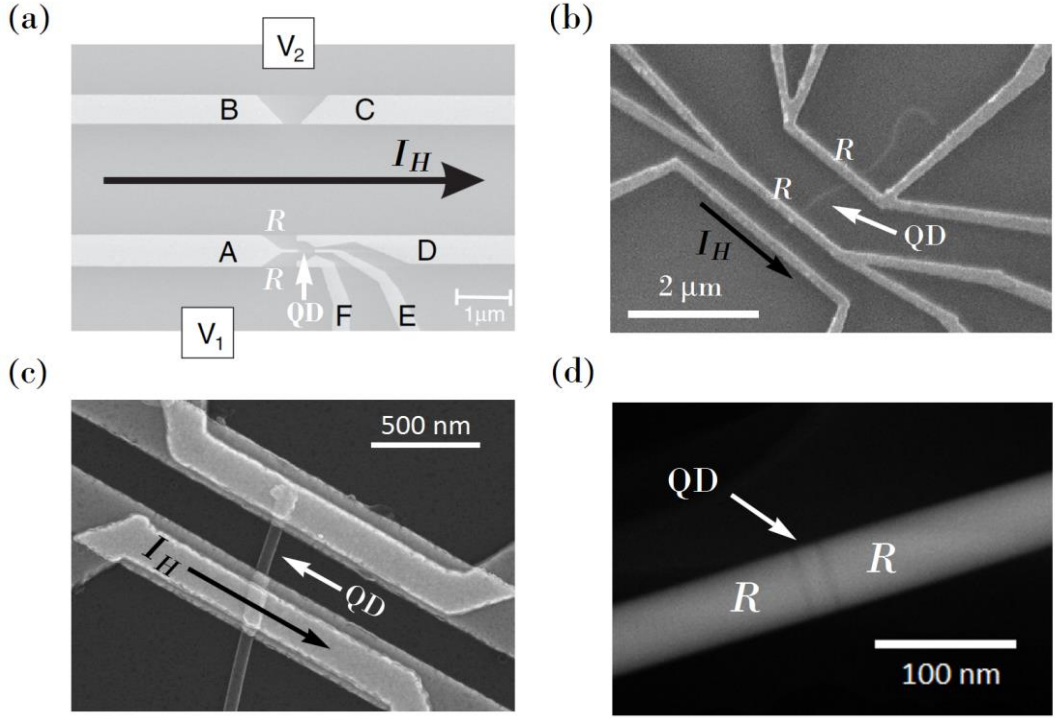

Figure 2: In the images black arrows indicate the heating elements (heated by a current $I_H$) used for thermal biasing. White arrows point to QD locations. Device parts serving as reservoirs are labeled with white "R" (a) Scanning electron microscope (SEM) image of a QD thermoelectric device defined in a 2DEG (adapted with permission from [25], copyrighted by American Physical Society). B and C can define a quantum point contact, if needed. A, D and A, F create tunnel-junctions, E is used as a plunger gate. $V_{th}$ is measured as a difference between potentials $V_1$ and $V_2$. (b) SEM image of a single-walled carbon nanotube QD thermoelectric device. Two leads contacting the nanotube are used for measuring $V_{th}$ (adapted with permission from [26], copyrighted by American Physical Society). Tunnel-junctions are created by Schottky barriers at the lead-nanotube interface. A thermal bias $\Delta T$ is created by running a heating current $I_H$ through an additional electrode in proximity of one of the contacting leads. Temperatures of both leads contacting the QD are calibrated by measuring the resistances of lead segments in four-point geometry. (c) SEM image of a thermoelectric device in which a pair of InP barriers is used to define an InAs QD in an InAs nanowire, similar to the one seen up close in (d). The nanowire segment containing a QD is located between contacting leads, which are covered in an insulator for decoupling heater electrodes placed on top. (d) Transmission electron microscope (TEM) image of an InAs nanowire containing two, approximately 4 nm thin, InP segments that define a QD.

External gates can be used to tune the QD resonance energies relative to the electrochemical potentials in the reservoirs. 2DEG-based devices typically use finger (plunger) gates (see Fig 1a). In nanowires and nanotubes finger gates are also often used [27, 28, 29], however, in case only one gate is needed, it is practical to use a global back gate instead, in which case the gate potential is applied directly to the whole substrate beneath the oxide upon which the device is made (see, for example, [24, 30, 31]).

## Thermal bias

Because of the sub-micrometer size of QDs, the study of their thermoelectric properties requires very large temperature gradients: a thermal bias of $\Delta T = 1$ K often corresponds to local gradients of $10^6$ K/m or more. The control of $\Delta T$ thus requires highly localized dissipation of energy and prevention of too much parasitic heating in the other reservoir. This is typically done by running a current through a small electrical component which dissipates Joule heat.

Historically the first method used for thermal biasing of QDs was to run a current through one of the electron reservoirs [5, 12, 13, 32, 33]. This was achieved by introducing two additional terminals to the 2DEG electron reservoirs, each of which could be electrically biased. In this method, the thermal and electrical biasing circuits are coupled and it is necessary to carefully balance the potentials applied to the heater segment so that the electric bias across the QD is unchanged. This method is often preferred due to a relatively simple fabrication process, particularly for 2DEG-based thermoelectric QD devices. However, because of the need for careful balancing for each setting of the heater current, this method typically does not allow simple tuning of the applied $\Delta T$. The biggest advantage of this heating method is its locality: the heat is dissipated directly in the reservoir itself.

In order to decouple electrical and thermal biasing circuits, the two circuits need to be electrically separated. This approach is used in studies of nanotubes and nanowires located on substrate oxides. So-called side-heating elements (see Fig 1b) can then be placed in the proximity of one of the tube or wire ends [26, 34, 35]. It essentially creates a temperature gradient along the substrate oxide and thus along the tube and wire. The side heater can typically be fabricated in the same process step as the metal contacts to the QD, making it very convenient. A disadvantage is that a relatively large amount of heating is necessary to create a limited temperature gradient, because much heat is dissipated into the substrate. This leads to global heating of the device and makes it more difficult to perform low temperature thermoelectric measurements.

A more sophisticated, yet more efficient method is covering contacting leads by an electrically insulating layer and running heating elements directly on top. Such top heaters allow larger $\Delta T$ with considerably less device heating, thus opening the door for truly low-temperature ($T < 1$ K) thermoelectric experiments on nanowires and nanotubes with conveniently tunable $\Delta T$ [36, 37]. A disadvantage of this method is that it involves several additional processing steps, like deposition of an insulating oxide layer and alignment and deposition of heaters, which considerably prolong the fabrication procedure.

**Thermometry**

Quantitative measurements of the Seebeck coefficient require precise knowledge of $\Delta T$, implying that it is necessary to measure the temperatures of the reservoirs with a precision much higher than $\Delta T$. This is challenging for QDs because they are sub-micrometer sized objects such that temperature measurements should in principle probe temperatures even more locally. In practice, the leads are good thermal conductors which have relatively homogeneous temperature distributions, allowing temperatures to be measured less locally.

In some 2DEG thermoelectric QD devices [13, 25] $\Delta T$ has been calibrated by measurements of the thermovoltage over a quantum point contact [38] connected to the heated reservoir. Unfortunately, similar to the case for QDs, the $V_{th}$ signals of quantum point contacts can be rather complex, which makes interpretation difficult. Other 2DEG experiments [5, 39] have calibrated temperature by characterizing Shubnikov-de-Haas resistance oscillations [40, 41] of the heated reservoir as a function of magnetic field.

In nanowire and nanotube QD thermoelectric devices, a convenient choice has been resistive thermometry. It is based on the temperature dependence of the resistance of the leads, which is typically measured in a four-point geometry. This method is simple and has a wide temperature applicability range, but in many cases the resistance loses its temperature dependence below 10-20 K. This is presumably because of the electron-phonon scattering, which decreases in strength with decreasing $T$, and thus becomes unimportant compared to impurity scattering according to Matthiessen's rule. Another approach, employed to measure $\Delta T$ across a QD embedded into a nanowire, is to use the QD itself as a thermometer, by comparing the thermocurrent, $I_{th}$, to the known conductance spectrum at a known temperature [42, 43]. This method relies on the behavior of the QD being consistent with theory which is discussed in latter sections.

Calibration by superconductor-insulator-metal junctions has been used for characterizing QD nano-cooler performance [44], but has, to our knowledge, not yet been used for studies of thermoelectric properties of QDs, and is generally limited to temperatures below critical temperatures of superconductors, for example 1.2 K for Al.

**Thermoelectric measurements**

Many thermoelectric experiments on QDs use AC lock-in techniques, which are based on the ability of lock-in amplifiers to detect small signals with chosen frequencies in otherwise noisy signals. One typically uses a sinusoidally varying heating current with frequency $f$. Ideally no DC offset in heating current is present so that the periodic modulation results in positive and negative current for exactly half of the period each. Because heating effects are the same for both current directions, maximal heating power is delivered twice per period and thermoelectric response is best detected at a frequency of $2f$.

This detection method is suitable for measuring the $V_{th}$ response to relatively small $\Delta T$. In principle it should minimize any effects from drive frequency $f$ or, in fact, any other frequency including zero (static signals). An important drawback of this method is that the resulting $V_{th}$ response of a QD is not sinusoidal in time, meaning that the thermoelectric response signal to the heating current is distributed also among other frequency components, which in principle should all be accounted for during quantitative analysis. Lock-in measurements are thus more complicated to use for quantitative analyses and can be particularly troublesome if the thermoelectric response to the heating current is somewhat nonlinear.

Such problems can be avoided by DC measurement techniques for which the thermoelectric response is read out directly, making them suitable for $\Delta T$ dependence characterization where signals become stronger and signal-to-noise ratio is better. In certain cases, however, current-voltage rectifying behavior of the QD itself can lead to a stable DC signal that originates purely from rectification of noise in the reservoirs [45], leading to a $V_G$ dependence of signals similar to a typical thermoelectric response.

## 3. Physical picture

The thermovoltage of a QD has a characteristic oscillatory behavior as a function of gate potential. In this section we provide an intuitive explanation for this behavior.

In traditional conductance measurements without thermal bias, the resonant level spectrum of QDs manifests itself as a series of peaks in differential conductivity $g = dI/dV$ as a function of a gate potential, $V_G$, where $dI$ is a change in current in response to a small electrical bias $dV$ applied to reservoirs. This is illustrated in Fig. 3a for the case of larger electrical bias $V$. In simple terms one can think of $V$ opening a window $eV$ in energy within which populations of electronic states across a QD are not balanced. Whenever the QD's resonance energy overlaps with the bias window, a current $I$ can flow.

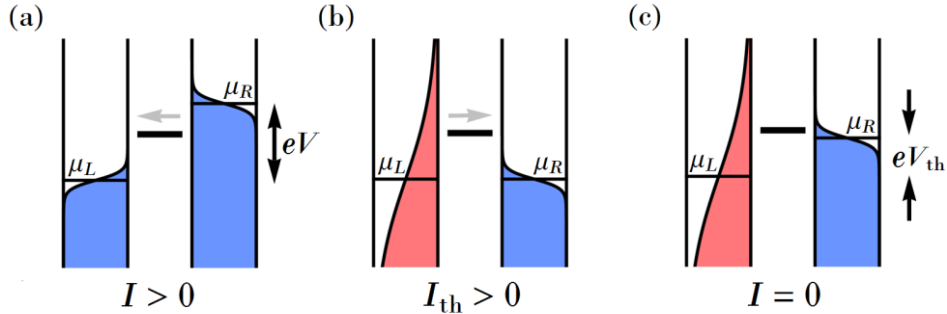

Figure 3: Illustrations of a resonance energy of a QD (black line) with respect to electrochemical potentials $\mu_L$ and $\mu_R$ of the reservoirs. Vertical axes represent electron energy. Colored areas represent Femi-Dirac distributions of electrons in reservoirs: blue indicates the temperature $T$, red indicates the higher temperature $T + \Delta T$. Gray arrows on top of resonances indicate electron current direction. Three cases: a) electrical bias giving rise to current $I$, b) thermal bias giving rise to thermocurrent $I_{th}$ (here shown at short-circuit condition), and c) thermal bias giving rise to thermovoltage $V_{th}$.

A thermal bias $\Delta T$ is an alternative source of misbalance of electron populations. Figure 3b illustrates how a $I_{th}$ can flow when a QD resonance is within the range of energies in which populations are unbalanced due to different temperatures.

If the circuit connecting the leads is opened, $I_{th}$ initially keeps transferring electrons from one side to the other. Gradually the potential difference between reservoirs is rising due to net charge accumulation. The process continues until the potential difference balances populations of electrons at the resonance energy and $I_{th}$ vanishes (Fig. 3c). This steady-state potential difference at open-circuit condition is called a thermovoltage, $V_{th}$. Because the electron populations are oppositely imbalanced below and above the electrochemical potentials of the reservoirs, the sign of $V_{th}$ and the direction of $I_{th}$ depend on the resonance energy position (relative to $\mu_L$ and $\mu_R$), which is tunable by $V_G$.

The above explanation considers only one resonance energy, whereas typically there are several that can all contribute as parallel transport channels. However, when $k_BT$ and $k_B(T+\Delta T)$ are much smaller than $\Delta\varepsilon$, the majority of transported electrons are carried by the resonance closest to $\mu_{L,R}$. By varying $V_G$, resonances can be thermoelectrically probed one by one. The overall dependence of $V_{th}$ on $V_G$ has the same periodicity as peaks in $g$, meaning that it changes sign twice per period. One of the two sign changes coincides with peaks in $g$, which occur when resonances cross $\mu_{L,R}$.

The first theoretical formulation explaining the $V_{th}$ behavior in QDs was given by Beenakker *et al.* [4] using the single-electron tunneling approximation (SETA) for thermoelectric transport, meaning that electron tunneling is treated in leading order perturbation theory, neglecting the finite width of the resonances as well as co-tunneling and other higher order tunneling processes. For small $\Delta T$ in the limit of classical Coulomb blockade ($\Delta E \ll k_BT \ll E_C$), this theory predicts a distinct sawtooth-shaped dependence of $V_{th}$ on $V_G$ (Fig. 4) The $V_{th}$ amplitude is predicted to change between $\pm V_0 = \pm e\Delta T/(4CT)$ with a sharp jump between positive and negative extremes precisely between sequential resonances and the linear change in $V_{th}$ for the rest of the $V_G$ range (Fig. 4).

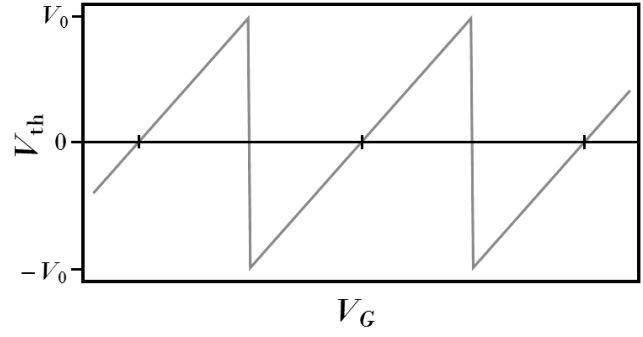

Figure 4: Sketch of the theoretically predicted $V_{th}$ dependency on $V_G$ in SETA in the limit of classical Coulomb blockade ($\Delta E \ll k_B T \ll E_C$) and in the linear $\Delta T$ response regime (see Ref.[4]). Ticks on $V_G$ axis correspond to positions of peaks in $g$ (Coulomb peaks). The amplitude of $V_{th}$ changes between $\pm V_0 = \pm e\Delta T/(4CT)$ as a function of $V_G$ in a sawtooth manner.

In addition to the SETA used by Beenakker, Landauer-type approaches using transmission formalism have often been used in comparisons with thermoelectric experiments. Such approaches take into account the finite resonance width, which is usually obtained by fitting $g$-measurements, but are not able to treat interactions beyond a mean-field approximation. As we will comment on below, Landauer-type approaches have nonetheless been used to qualitatively explain experiments on Coulomb-blockaded QDs with reasonably good success. A large number of more sophisticated theoretical approaches exist which include both local interaction effects on the QD (such as Coulomb blockade), higher order tunneling and resonance broadening, but we do not attempt a review of this wide theory field here.

The so-called Mott's relation for thermovoltage is sometimes used to obtain an estimate of $V_{th}$ or $S$ based on a conductance measurement alone [13]. It states that if $g$ varies slowly over an energy range $k_B T$, then the linear response to $\Delta T$ should give $S = V_{th}/\Delta T = -\pi^2 k_B^2 T g'/(3eg)$, where $g' = dg/d\varepsilon$. This relation is most appropriate for metallic systems and also has been somewhat successful in predicting $V_{th}$ for degenerate semiconductor QDs in the regime where $k_B T < \hbar\Gamma$ [13, 25]. However, it is not applicable for QDs in the limit $\hbar\Gamma \ll k_B T \ll \Delta\varepsilon$ where Beenakker's theory predicts a perfect sawtooth-like gate dependence of $V_{th}$.

# 4. Experiments on thermovoltage and thermocurrent in quantum dots

Staring *et al.* [12] were the first to perform $V_{th}$ measurements on a QD and to confirm the expected oscillatory behavior described above. The device was defined by surface gates in a 2DEG and a $\Delta T$ was applied by running a current through one of the reservoirs. The authors observed $V_G$ dependent oscillations in $V_{th}$ that had the same period as peaks in $g$ (Fig. 5). The shape of the $V_{th}$ signal showed a sawtooth-like behavior with positive slope at the $g$ peak positions and a sharper negative slope in between peaks. The results were in good qualitative agreement with SETA [4], but no quantitative agreement with the theory was achieved. The best qualitative fit to their results was found for $T$ = 0.23 K, which was almost five times higher than the lattice temperature $T_l$ = 50 mK of the device. The study's authors suggested that this deviation might have originated from a finite resonance width, $\hbar\Gamma > k_B T$, which is neglected in SETA, and is also expected to reduce $V_{th}$, similarly as higher $T$. A direct comparison of the measured $V_{th}$ with theory was not possible since reliable measurements of $\Delta T$ were not available.

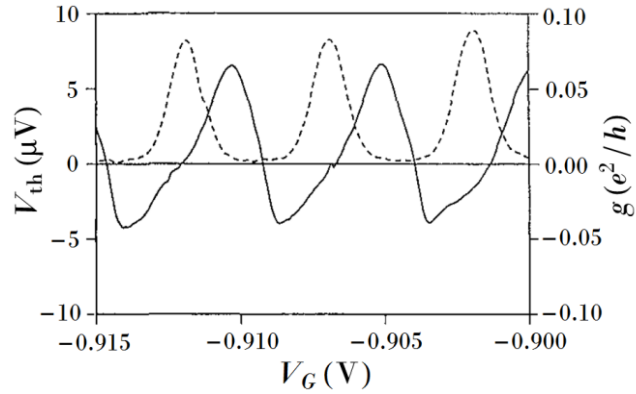

Figure 5: Measured $g$ (dashed line) and $V_{th}$ (solid line) as functions of $V_G$ for a 2DEG-based QD. Figure adapted with permission from [12], copyrighted by EDP Sciences.

In similar experiments on a 2DEG device, Dzurak *et al.* [13] independently found similar behavior in the regime $k_B T < \hbar\Gamma < E_C$ where the SETA is not expected to be applicable. Because the authors succeeded in calibrating $\Delta T$ by a QPC, a quantitative comparison with theory was possible. Results showed quantitatively similar behavior to what was predicted by Mott's relation even when $\Delta T/T > 1$, whereas the SETA gave an overestimate of $V_{th}$ by about an order of magnitude.

In a follow-up study a few years later, Dzurak *et al.* investigated a QD in a wider energy range for which $\Delta T$ was estimated based on the behavior of SdH oscillations in the heated channel [5]. Their results (see Fig. 6a) showed a series of peaks in $g$ with amplitude and with widths that generally increases with $V_G$. The measured $V_{th}$ oscillations had a period identical to the one of $g$ peaks (Fig. 6b), just as seen in previous studies [12, 13], but also decayed at more negative $V_G$. However, the authors also pointed out that within this $V_G$ range the QD resistance exceeded 200 MΩ and any small current leakages could have potentially shortened out the $V_{th}$ signal.

A new experimental discovery in this experiment was a fine structure on top of the large $V_{th}$ oscillation period (Fig. 6c) that was previously predicted [4] for energetically well separated excited states of the QD.

Dzurak *et al.* estimated that, given $T$ = 50 mK and $\Delta T$ = 170 mK in their device, the measured $V_{th}$ was about two orders of magnitude smaller than predicted by the SETA [4]. This deviation was again attributed to the fact that the SETA does not take the resonance broadening into account. However, the authors also found that a Landauer-type equation with the same parameters overestimated $V_{th}$ by only a factor of 2.5 and the best fit was found for $\Delta T$ = 100 mK, that is, almost a factor of two smaller than the measured $\Delta T$.

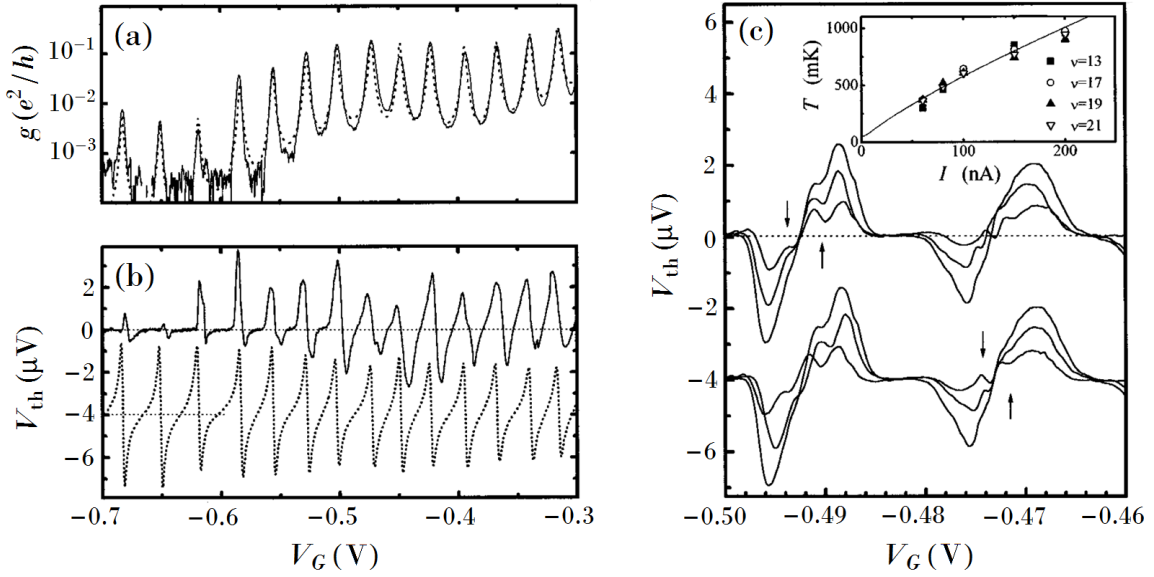

Figure 6: Figures from Ref. [5]. (a) $g$ as a function of $V_G$. (b) $V_{th}$ as a function of $V_G$. In (a) and (b), solid lines represent experimental data. The dotted line shows Landauer-type simulation results with the following parameters: gate coupling constant $\alpha = 0.54$, $\Delta T = 100$ mK, $T = 50$ mK (unheated), $E_C = 0.75$ meV. In (b) $V_{th}$ simulation results are off-set by 4 µV. (c) $V_{th}$ measurement results for 30, 40 and 50 nA heating current amplitude as a function of gate potential. The two sets of curves, offset by 4 µV, are the same data, but shifted in gate potential to cross zero at the same potential value marked by arrows. Inset: $T + \Delta T$ calibration results in heated reservoir. Figures adapted with permission from [5], copyrighted by American Physical Society.

A study on the behavior of the Seebeck coefficient $S = V_{th}/\Delta T$ as a function of $T$ was published by Small et al. [26]. Unlike earlier experiments, this study used a single-walled carbon nanotube in which a QD was defined by Schottky contacts to metallic leads. This experiment pioneered thermal biasing using the side heating architecture (Fig. 2b), and $\Delta T$ was calibrated by resistive thermometry in the leads. A conceptually similar experimental device was also investigated (without probing $T$) in another study by Llaguno et al. [35] and published shortly after.

Results by Small et al. are presented in Fig. 7. Owing to a large average value of $E_C = 6$ meV, the resolution of $S$ oscillations was possible at temperatures as high as $T = 30$ K. Like in the earlier work, the oscillations in $S$ correlated with peaks in $g$, but were more irregular. Instead of discussing the fine shape of the $S$ signal, the average root-mean-squared (rms) amplitude of $S$ oscillations was analyzed as a function of $T$. $S_{rms}$ was found to be several times smaller than what SETA would predict. However, as can be seen in the inset of Fig. 7, $S_{rms}$ was observed to be proportional to $1/T$, as predicted by Beenakker et al. [4].

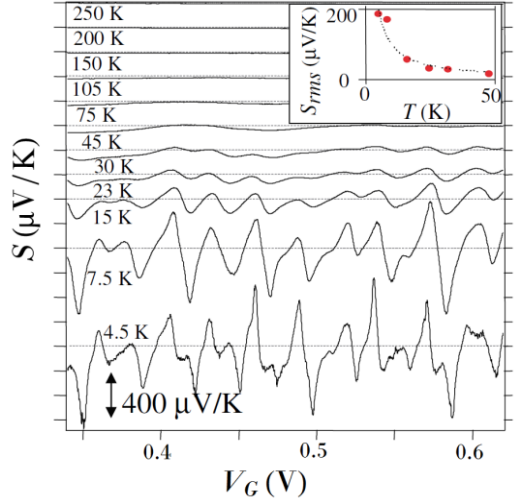

Figure 7: S = $V_{th}/\Delta T$ as a function of $V_G$, at different $T$ ranging from 4.5 K to 250 K. Curves are off-set by margins indicated by dotted lines. Inset: Root-mean-squared of $V_{th}/\Delta T$ amplitude depending on temperature. The dotted line indicates a fit that scales as $1/T$. Figure adapted with permission from [26], copyrighted by American Physical Society.

More detailed investigations of the fine $V_{th}$ oscillation dependence on $T$ were carried out by Scheibner et al. [5] a few years later. One of their two 2DEG-based devices showed that the magnitude of $V_{th}$ as a function of $T$ evolves differently at different $V_G$ (see Fig. 8). More precisely, while the $V_{th}$ signal was observed to decay with increasing $T$ at its peaks, at other $V_G$ values $V_{th}$ showed an enhancement with increasing $T$.

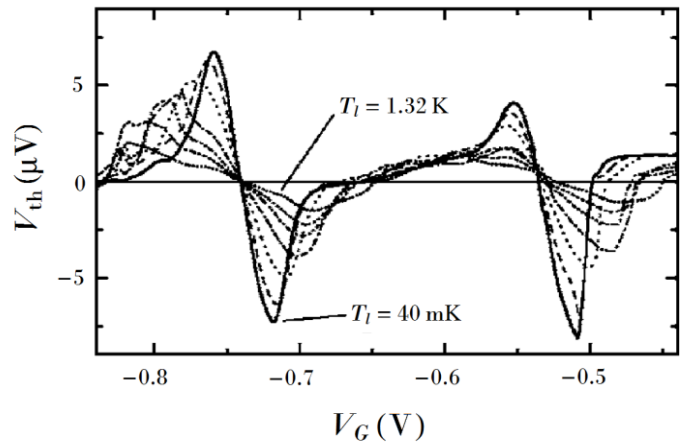

Figure 8: Thermovoltage, $V_{th}$, as a function of $V_G$, for $T_l$ = 39, 66, 158, 257, 425, 1040 and 1319 mK. Figure adapted with permission from [6], copyrighted by American Physical Society.

Scheibner et al. used two theoretical approaches for comparison with their data: the SETA [4] and an approach accounting for co-tunneling processes [46]. Neither of the approaches was found to correctly describe the $V_{th}$ peak positions, their height or their evolution with $T$ over the whole range. Generally, agreement was better for higher $T_l$ where both models gave similar results. These results showed that the SETA does not correctly describe thermoelectric transport at low $T$ and, depending on $\hbar\Gamma/k_BT$, resonance broadening and higher order tunneling processes can strongly influence the behavior of $V_{th}$. Precise values of $\hbar\Gamma/k_BT$ at which this behavior becomes important, however, could not be obtained from the results.

One of the most recent experimental studies analyzing the dependence of $V_{th}$ on $T$ was published in 2012 by Fahlvik Svensson et al. [32] using QDs defined by a pair of InP barriers in InAs semiconductor nanowires. $\Delta T$ was created by running a heating current through one of the contacting metal leads, similar to the 2DEG experiments described above. Similar $V_{th}$ evolution with $T$, as demonstrated by Scheibner et al. [5], was found. As can be seen in Fig. 9, at lower $T$, $V_{th}$ oscillations were sharp and localized, whereas at higher $T$ the $V_{th}$ peaks became broader and showed more sawtooth-like behavior. Fahlvik Svensson *et al.* compared their data with Landauer-like simulations, and found that the evolution of the line shape as a function of $T$ could be understood in terms of an evolution of the energy-scale ratios $\hbar\Gamma/k_BT$ and $\Delta\varepsilon/k_BT$. In fact, after taking RC constants of measurement circuits into account, the calculated $V_{th}$ magnitude was shown to be off by roughly a factor of two all throughout the temperature range, while giving excellent qualitative agreement. This suggested that in certain cases increasing $T$ alone can be enough to account for $V_{th}$ line shape evolution. It also illustrated the need to properly take into account the finite input impedances and damping time constants that are present in measurement setups: because the resistance of a QD varies strongly between on- and off-resonance conditions, damping effects become $V_G$ dependent and thus not only affect the amplitude but also the gate dependence (shape) of the observed quantities.

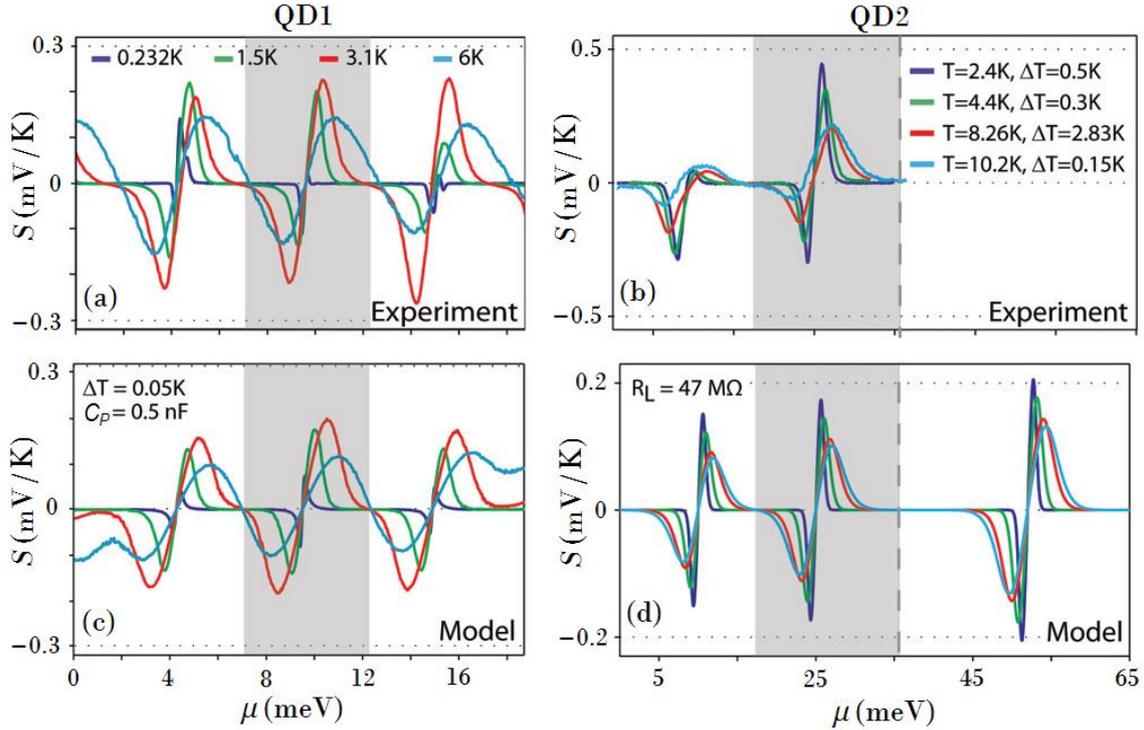

Figure 9: (a) and (b) are experimentally measured $S$ as a function of the QD electrochemical potential energy $\mu$ (proportional to $V_G$) for two nanowire QD devices [32]. (c) and (d) are corresponding simulation results taking RC effects into account. For (a) cryostat temperatures $T$ are indicated in the figure and $\Delta T$ are estimated by finite element simulations. For (b) cryostat temperatures $T$ and $\Delta T$, both indicated in the figure, where $\Delta T$ areestimated by quantum dot thermometry [42, 43]. (c) Simulation results based on Landauer-type approach which also accounts for RC damping of ac signal in measurement circuit with 0.5 nF parasitic capacitance to ground, $C_P$. (d) Simulation results based on Landauer-type approach which also accounts for reduction of voltage in measurement circuit due to an RC filter in parallel with the QD device. Figure adapted with permission from [32].

The studies reviewed so far have mainly been interpreted within a linear response picture. Generally, however, neither $V_{th}$ nor $I_{th}$ in QDs scale linearly with $\Delta T$ in a wide range. Essentially the origin of these nonlinearities is the nonlinear character of the Fermi-Dirac distribution and sharply localized resonances in energy. In fact, already in their report of the very first experimental study of $V_{th}$ in a QD, Staring *et al.* [12] noticed nonlinear behavior of $V_{th}$ as a function of $\Delta T$. However, apart from this initial observation, nonlinear thermoelectric effects in QDs have received limited experimental attention [33, 37, 47]. In the work published by Fahlvik Svensson *et al.* [33] three QDs in semiconductor nanowires were investigated. Theoretical models were unable to qualitatively explain the strongly nonlinear behavior of $I_{th}$ and $V_{th}$, including sign changes as a function of $\Delta T$, unless some kind of temperature dependence of the resonance spectrum was assumed. Even though signs of such resonance renormalization were seen, the side-heating geometry did not allow discrimination between renormalization effects appearing due to $\Delta T$ or due to the associated parasitic heating of the device.

Another version of the experiment was carried out by Svilans *et al.* [37] using a similar device to that shown in Fig. 2c. The device featured top-heaters [36] which reduced overall heating compared to the side heaters used in [33], while being electrically independent of the electric measurement circuit. Figure 10a shows their measurement results on $I_{th}$ as a function of $I_H$ (roughly proportional to $\Delta T$). Different curves present different $V_G$ settings, for all of which the same two resonances have $\mu_L$ and $\mu_R$ between them. $I_{th}$ increases in magnitude, saturates, and eventually decays with increasing $I_H$. This behavior can be explained with the help of Fig. 10b, which illustrates distributions of thermally excited electrons relative to resonance energies for increasing $\Delta T$. At first, $\Delta T$ gives rise to $I_{th}$ via the resonance closest to $\mu_{L,R}$. As $\Delta T$ increases, $I_{th}$ increases super-linearly due to the exponential nature of Fermi-Dirac distribution tails. At a certain $\Delta T$, electrons start being transported through the other resonance which contributes to current in the opposite direction. Depending on how both resonance energies are positioned with respect to $\mu_{L,R}$ the onset, saturation and decay of $I_{th}$ happened at different $\Delta T$ values as well as with different polarities and amplitudes.

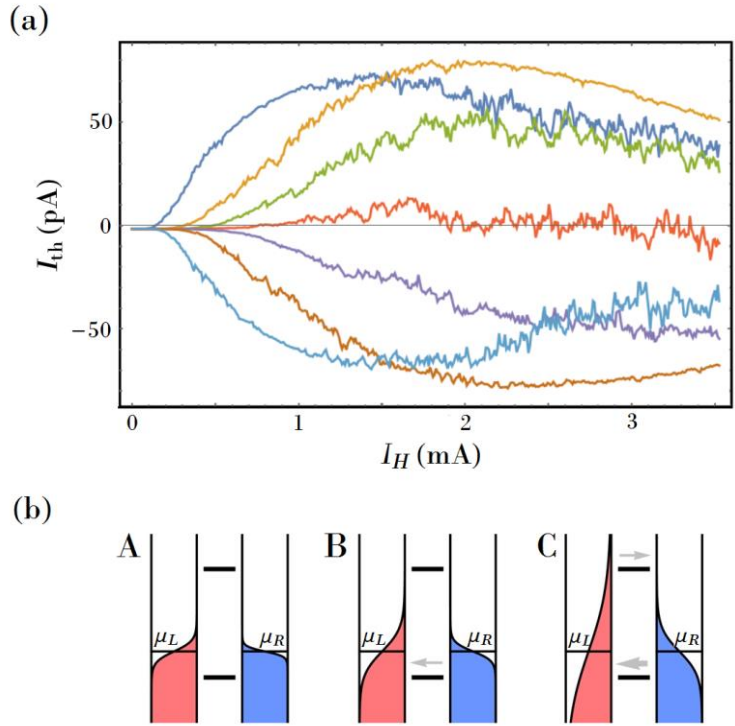

Figure 10: (a) Experimental results on thermocurrent, $I_{th}$, depending on heating current, $I_H$, for 7 approximately equidistant resonance energy positions relative to electrochemical potentials of the leads, $\mu_L$ and $\mu_R$. Figure adapted with permission form [37]. (b) Sketches of Fermi-Dirac distributions of electrons in the hot (red) and cold (blue) electron reservoirs relative to resonance energies for increasing thermal bias from A to C (similar to Fig. 3). Gray arrows on top of resonances indicate electron current direction.

Svilans *et al.* compared their results to a model based on SETA [4, 15]. Quantitative comparison between theory and experiment was not possible because neither $T$ nor $\Delta T$ were measured in the experiment. Nevertheless, by assuming linearly increasing $T$ and $\Delta T$ with heating current, the authors were able to reproduce qualitative behavior of $I_{th}$ in great detail over a $k_B \Delta T$ range that was supposedly a good fraction of $E_C = 2$ meV. This result was also in agreement with an earlier theoretical prediction by Sanchez *et al.* [48] who also considered two resonances, but used a theoretical approach that took widths $\hbar \Gamma$ into account.

## 5. Discussion

**Applicability of theories.** Much discussion throughout the studies reviewed here has been dedicated to the applicability of theoretical models. It is known that $\hbar\Gamma \ll k_BT$ is a strong condition for the SETA to work. For differential conductivity measurements a typically used condition for validity of SETA is $g \ll e^2/h$. However, despite this condition being met in most studies reviewed here, quantitative agreement with theory for amplitude and $V_G$ dependence of $V_{th}$ has not been found.

In fact, $V_{th}$ is very sensitive to the magnitude of $\hbar\Gamma$ even when $\hbar\Gamma \ll k_BT$. This can be seen by looking at Landauer-type simulation results which converge towards the ones carried out in SETA exponentially slowly as a function of $\hbar\Gamma$ [33]. In this limit, however, widths of $g$ peaks cannot be used for estimating $\Gamma$ directly because they themselves are broadened by $T$, and the amplitude of the current gives access only to the related quantity $\gamma$. Therefore it is difficult to use conductance measurements for confirming that a non-negligible $\Gamma$ is the reason for the observed disagreement, unless $T$ can be farther lowered to $\hbar\Gamma \approx k_BT$.

In the opposite limit, when $k_BT < \hbar\Gamma$, Mott's relation has been relatively successful in predicting $V_{th}$. However, this is also the limit in which electron correlation effects have been introducing systematic deviations from Mott's law [25]. This has been experimentally studied in more detail in two PhD theses [49, 50], however, because thermoelectric behavior in highly correlated QDs is even more complex, comparison with theory is even more difficult.

**Excited states and asymmetries.** Apart from the amplitude of $V_{th}$, its dependence on $V_G$ is also very often observed to disagree with theory. The typical $V_{th}$ oscillation "shape" found in measurements is frequently bent, modulated, or has strongly asymmetric positive and negative amplitude parts. Dzurak *et al*. [5] demonstrated that one source of such deviations can be excited QD states, as predicted by the SETA, although their measured $V_{th}$ amplitude did not agree with that predicted by the SETA, most likely because $\hbar\Gamma \ll k_BT$ was not fulfilled in the experiment. However, there is still very little experimental support for the full extent of effects due to excited states or higher order tunneling processes on $V_{th}$ (and $I_{th}$).

Asymmetries in QD tunnel- and capacitive couplings can also have an effect on the thermoelectric behavior of QDs. It is well established that in the nonlinear electrical bias regime asymmetric capacitive and tunnel-couplings in QDs can lead to asymmetries in the current-voltage (*I-V*) characteristics (rectification). In fact, $G$ needs to be mapped in the nonlinear regime in order to obtain relevant information about excited states of a QD or asymmetries in couplings. It is an interesting question how asymmetries in the couplings affect thermoelectric response of QDs in the nonlinear regime. In the limit $k_B\Delta T \ll E_C$ asymmetric tunnel-couplings have only been briefly discussed in relation to thermoelectric measurements by Scheibner *et al*. [51], who qualitatively explained a pronounced asymmetry between positive and negative $V_{th}$ amplitudes by assuming two nearly degenerate QD states, the ground state of which was of blocking nature due to a much weaker tunnel-coupling to one of the reservoirs.

**Thermocurrent and thermovoltage.** In the linear response regime, $I_{th}$ relates to $V_{th}$ via the differential conductivity $g$ as $I_{th} = gV_{th}$. Because $g$ in QDs has a strong $V_G$ dependency, the behaviors of $I_{th}$ and $V_{th}$ as functions of gate are qualitatively different. This also means that $I_{th}$ measurements probe thermoelectric properties in a slightly different way than $V_{th}$, and that $I_{th}$ is not a one to one substitute for $V_{th}$ – the two measurements complement each other.

In general, the biggest conceptual difference, when measuring $I_{th}$ as compared to $V_{th}$, is that the potential difference across a QD can be kept low or ideally at zero, such that any effects that may arise due to a finite bias across the QD can be avoided. One could say that $I_{th}$ is a quantitative measure of the net thermoelectrically transported charge per unit time, whereas $V_{th}$ is more related to the average energy of the transported charges. One particularly interesting direction related to the nonlinear regime is the use of QDs for their thermal-to-electric energy conversion properties. Because of the generally unknown relation between $V_{th}$ and $I_{th}$ in the nonlinear response regime, thermoelectric power cannot simply be obtained from the power factor $gS^2$ and must be measured directly instead.

# 6. Summary and conclusions

In conclusion, we reviewed basic concepts of single quantum dot thermoelectric devices, gave an overview of fundamentally important thermoelectrics experiments on quantum dots, and discussed previously little discussed aspects of the field. We found that the commonly used single-electron tunneling approximation has not been successful in explaining the magnitude of thermovoltage in quantum dots. Instead, considering electron transmission properties of quantum dots empirically has given more consistent qualitative agreement yet still overestimated the thermovoltage amplitude. Future work is expected to include studies of the thermoelectrically produced power of quantum dots and studies of thermovoltage and thermocurrent of highly asymmetric quantum dots in the nonlinear thermal bias regime.

**Acknowledgements.** We thank the authors of References [5, 6, 12, 25, 26, 32], for permission to adapt material in Figures 2, 5, 6, 7, 8 and 9. We also thank S. Lehmann for the TEM image and A. M. Burke for the SEM image in Fig. 2c. This work was supported by the People Programme (Marie Curie Actions) of the European Union's Seventh Framework Programme (FP7-People-2013-ITN) under REA grant agreement n°608153, by the Swedish Energy Agency (project P38331-1), by the Swedish Research Council (project 621-2012-5122) and by NanoLund.

# References

[1] L.P. Kouwenhoven et al., Electron transport in quantum dots, Kluwer Series **E345** (1997) Proceedings of the NATO Advanced Study Institute on Mesoscopic Electron Transport 105

[2] K. K. Likharev, Single-electron devices and their applications, Proceedings of the IEEE **87** (1999) 4

[3] D.M. Rowe, Thermoelectrics Handbook: Macro to Nano, CRC Press, Boka Raton, 2006

[4] C. W. J. Beenakker and A. A. M. Staring, Theory of the thermopower of a quantum dot, Phys. Rev. B **46** (2002) 9667

[5] A. S. Dzurak et al., Thermoelectric signature of the excitation spectrum of a quantum dot, Phys. Rev. B **75** (1997) 041301

[6] R. Scheibner et al., Sequential and cotunneling behavior in the temperature-dependent thermopower of few-electron quantum dots, Phys. Rev. B **75** (2007) 041301R

[7] G. D. Mahan and J. O. Sofo, The best Thermoelectric, PNAS **93** (1996) 7436

[8] B. Kubala, J. König, and J. Pekola, Violation of the Wiedemann-Franz law in a single-electron transistor, Phys. Rev. Lett. **100** (2008) 066801

[9] T. E. Humphrey et al., Reversible quantum brownian heat engines for electrons, Phys. Rev. Lett. **89** (2002) 116801

[10] N. Nakpathomkun, H. Q. Xu, and H. Linke, Thermoelectric efficiency at maximum power in low-dimensional systems, Phys. Rev. B **82** (2010) 235428

[11] H. Thierschmann et al., Thermoelectrics with Coulomb coupled quantum dots, C. R. Physique (2016)

[12] A. A. M. Staring et al., Coulomb-blockade oscillations in the thermopower of a quantum dot, Europhys. Lett. **22** (1993) 57 (1993)

[13] A. S. Dzurak et al., Observation of Coulomb blockade oscillations in the thermopower of a quantum dot, Solid State Commun. **87** (1993) 1145

[14] C. C. Escott, F. A. Zwanenburg and A. Morello, Resonant tunneling features in quantum dots, Nanotechnology **21** (2010) 274018

[15] C. W. J. Beenakker, Theory of Coulomb-blockade oscillations in the conductance of a quantum dot, Phys. Rev. B **44** (1991) 1646

[16] L. P. Kouwenhoven et al., Single electron charging effects in semiconductor quantum dots, Z. Phys. B - Condensed Matter **85** (1991) 367

[17] M. A. Reed et al., Observation of discrete electronic states in a zero-dimensional semiconductor nanostructure, Phys. Rev. Lett. **60** (1988) 535

[18] L. P. Kouwenhoven, D. G. Austing and S. Tarucha, Few-electron quantum dots, Rep. Prog. Phys. **64** (2001) 701

[19] E. Bonet, M. M. Deshmukh and D. C. Ralph, Solving rate equations for electron tunneling via discrete quantum states, Phys. Rev. B 65 (2002) 045317

[20] L. Pfeiffer et al., Electron mobilities exceeding 107 cm2/Vs in modulation-doped GaAs, Appl. Phys. Lett. **55** (1989) 1888

[21] H. A. Nilsson et al., InSb nanowire field-effect transistors and quantum-dot devices, IEEE J. Sel. Top. Quantum Electronics **17** (2011) 907

[22] M. Freitag et al., Controlled creation of a carbon nanotube diode by a scanned gate, Appl. Phys. Lett. **79** (2001) 3326

[23] M. T. Björk et al., One-dimensional steeplechase for electrons realized, Nano Letters **2** (2002) 87

[24] M. T. Björk et al., Few-electron quantum dots in nanowires, Nano Letters, **4** (2004) 1621

[25] R. Scheibner et al., Thermopower of a Kondo spin-correlated quantum dot, Phys. Rev. Lett. **95** (2005) 176602

[26] J. P. Small, K. M. Perez, and P. Kim, Modulation of thermoelectric power of individual carbon nanotubes, Phys. Rev. Lett. **91** (2003) 256801

[27] M. J. Biercuk et al., Gate-defined quantum dots on carbon nanotubes, Nano Lett. **5** (2005) 1267

[28] S. Roddaro et al., Spin states of holes in Ge/Si nanowire quantum dots, Phys. Rev. Lett. **101** (2008) 186802

[29] V. S. Pribiag et al., Electrical control of single hole spins in nanowire quantum dots, Nature Nanotechn. **8** (2013) 170

[30] S. Sapmaz et al., Quantum dots in carbon nanotubes, Semicond. Sci. Technol. **21** (2006) S52

[31] H. A. Nilsson et al., Giant, level-dependent g factors in InSb nanowire quantum dots, Nano Lett. **9** (2009) 3151

[32] S. Fahlvik Svensson et al., Lineshape of the thermopower of quantum dots, New J. Phys. **14** (2012) 033041

[33] S. Fahlvik Svensson et al., Nonlinear thermovoltage and thermocurrent in quantum dots, New J. Phys. **15** (2013) 105011

[34] J. P. Small and P. Kim, Thermopower measurements on individual single walled carbon nanotubes, Mic. Thermophys. Engin. **8** (2004) 1

[35] M. C. Llaguno, J. E. Fischer, and A. T. Johnson, Jr., Observation of thermopower oscillations in the Coulomb blockade regime in a semiconducting carbon nanotube, Nano. Lett. **4** (2004) 45

[36] J. G. Gluschke et al., Fully tunable, non-invasive thermal biasing of gated nanostructures suitable for low-temperature studies, Nanotechnology **25** (2014) 385704

[37] A. Svilans et al., Nonlinear thermoelectric response due to energy-dependent transport properties of a quantum dot, Physica E **82** (2016) 34

[38] H. van Houten et al., Thermo-electric properties of quantum point contacts, Semicond. Sci. Tech. **7** (1992) B215

[39] A. S. Dzurak et al., Thermopower measurements of semiconductor quantum dots, Physica B **281** (1998) 249

[40] T. Ando, Theory of quantum transport in a two-dimensional electron system under magnetic fields. IV. Oscillatory Conductivity, J. Phys. Soc. Jpn. **37** (1974) 1233

[41] R. Fletcher et al., Hot-electron temperatures of two-dimensional electron gases using both de Haas–Shubnikov oscillations and the electron-electron interaction effect, Phys. Rev. B **45** (1992) 6659

[42] E. A. Hoffmann et al., Quantum dot thermometry, Appl. Phys. Lett. **91** (2007) 252114

[43] E. A. Hoffmann et al., Measuring temperature gradients over nanometer length scales, Nano Lett. **9** (2009) 779

[44] A. V. Feshchenko, J. V. Koski, and J. P. Pekola, Experimental realization of a Coulomb blockade refrigerator, Phys. Rev. B **90** (2014) 201407R

[45] J. Weis et al., Single-electron tunnelling transistor as a current rectifier with potential-controlled current polarity, Semicond. Sci. Tech. **10** (1995) 877

[46] M. Turek and K. A. Matveev, Cotunneling thermopower of single electron transistors, Phys. Rev. B **65** (2002) 115332

[47] A. G. Pogosov et al., Coulomb blockade and the thermopower of a suspended quantum dot, JETP. Lett. **83** (2006) 122

[48] M. A. Sierra, D. Sanchez, Strongly nonlinear thermovoltage and heat dissipation in interacting quantum dots, Phys. Rev. B **90** (2014) 115313

[49] R. Scheibner Ph. D dissertation, Wurzburg, Germany (2007)

[50] H. Thierschmann Ph. D dissertation, Wurzburg, Germany (2014)

[51] R. Scheibner et al., Quantum dot as thermal rectifier, New Jour. Phys. **10** (2008) 083016